\documentclass[lettersize,journal]{IEEEtran}
\usepackage{amsmath,amsfonts}
\usepackage{algorithmic}
\usepackage{algorithm}
\usepackage{array}
\usepackage[caption=false,font=normalsize,labelfont=sf,textfont=sf]{subfig}
\usepackage{textcomp}
\usepackage{stfloats}
\usepackage{url}
\usepackage{bigstrut}
\usepackage{multirow}
\usepackage{balance}
\usepackage{verbatim}
\usepackage{graphicx}
\usepackage{cite}
\usepackage{amssymb}
\usepackage[backref]{hyperref}
\usepackage[dvipsnames]{xcolor}
\begin{document}
	
	\title{MGCR-Net:Multimodal Graph-Conditioned Vision-Language Reconstruction Network for Remote Sensing Change Detection}
	
	\author{Chengming~Wang, Guodong Fan, Jinjiang Li, Min Gan, ~\IEEEmembership{Senior Member,~IEEE}, \\ C. L. Philip Chen, ~\IEEEmembership{Life Fellow, IEEE}
		\thanks{This research was supported by the National Natural Science Foundation of China (62301105, 61772319, 62272281, 62002200, 62202268), Shandong Natural Science Foundation of China (ZR2020QF012 and ZR2021MF068), Yantai science and technology innovation development plan(2022JCYJ031).}
		\thanks{Chengming Wang, Guodong Fan and Jinjiang Li are with School of Computer Science and Technology,
			Shandong Technology and Business University, Yantai 264005, China}
		\thanks{Min Gan is with School of Computer Science and Technology,
			Qingdao University, Qingdao 266000, China}
		\thanks{C. L. Philip Chen is with the School of Computer Science and Engineering, South China University of Technology, Guangzhou 510641, China}
	}
	\markboth{Journal of \LaTeX\ Class Files,~Vol.~14, No.~8, August~2021}%
	{Shell \MakeLowercase{\textit{et al.}}: A Sample Article Using IEEEtran.cls for IEEE Journals}
	
	\IEEEpubid{0000--0000/00\$00.00~\copyright~2021 IEEE}
	
	\maketitle
	
	\begin{abstract}
		With the advancement of remote sensing satellite technology and the rapid progress of deep learning, remote sensing change detection (RSCD) has become a key technique for regional monitoring. Traditional change detection (CD) methods and deep learning-based approaches have made significant contributions to change analysis and detection, however, many outstanding methods still face limitations in the exploration and application of multimodal data. To address this, we propose the multimodal graph-conditioned vision-language reconstruction network (MGCR-Net) to further explore the semantic interaction capabilities of multimodal data. Multimodal large language models (MLLM) have attracted widespread attention for their outstanding performance in computer vision, particularly due to their powerful visual-language understanding and dialogic interaction capabilities. Specifically, we design a MLLM-based optimization strategy to generate multimodal textual data from the original CD images, which serve as textual input to MGCR. Visual and textual features are extracted through a dual encoder framework. For the first time in the RSCD task, we introduce a multimodal  graph-conditioned vision-language reconstruction mechanism, which is integrated with graph attention to construct a semantic graph-conditioned reconstruction module (SGCM), this module generates vision-language (VL) tokens through graph-based conditions and enables cross-dimensional interaction between visual and textual features via multihead attention. The reconstructed VL features are then deeply fused using the language vision transformer (LViT), achieving fine-grained feature alignment and high-level semantic interaction. Experimental results on four public datasets demonstrate that MGCR achieves superior performance compared to mainstream CD methods. Our code is available on https://github.com/cn-xvkong/MGCR.
	\end{abstract}
	
	
	\section{Introduction}
	\IEEEPARstart{R}{SCD} focuses on identifying changes in ground objects from bi-temporal images and has become an inevitable outcome of remote sensing advancements aimed at achieving regional monitoring. Specifically, this technology can be applied to land use statistics, disaster impact assessment, and forest cover mapping, among other tasks \cite{1,2,3}. In this work, we focus on binary CD to construct MGCR, where the prediction results are presented as binary classification maps. Traditional CD methods include pixel-level, object-level, and scene-level CD \cite{4,5,6}, which have demonstrated strong performance in manually labeled change analysis but require substantial resources and suffer from limited adaptability. Machine learning approaches learn from raw data and extract features to adaptively optimize model weights. Algorithms such as support vector machines, random forests, and k-means clustering \cite{7,8,9,10} have been employed to perform binary classification and detect changes between bi-temporal images from different perspectives. However, they still exhibit limitations in capturing deep semantic information.
	
	\IEEEpubidadjcol
	
	\begin{figure}[!t]
		\centering
		\includegraphics[width=3.5in]{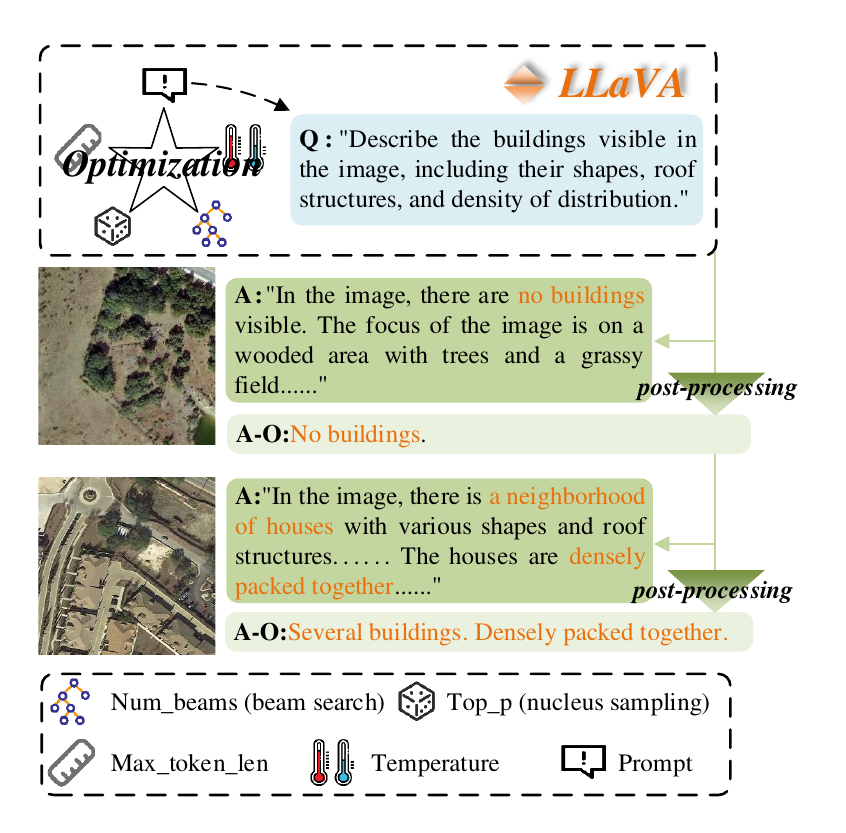}
		\caption{Illustrates our optimization strategy for LLaVA. By designing tailored prompts and tuning multiple parameters, we guide the model to generate text descriptions relevant to the input images. Additionally, we apply regularization-based pruning to refine and enhance the generated text, ensuring higher quality and better alignment with visual content.
		}
		\label{fig1}
		\vspace{-15pt}
	\end{figure}

	The development of deep learning methods has effectively addressed the limitations of traditional machine learning in CD tasks by significantly simplifying the complex feature engineering processes involved in conventional CD pipelines, while also reducing human intervention and computational costs. Convolutional neural networks (CNNs), with their strong local perception capabilities and parameter-sharing mechanisms, can efficiently extract local features from images, enabling accurate pixel-level classification and precise localization of change regions. However, CNNs are inherently limited in modeling long-range dependencies and struggle to capture global correlations between non-local features in images. The introduction of the transformer architecture \cite{11} offers an effective solution to this issue. Through the self-attention mechanism, transformers can model long-distance dependencies across the entire image, thereby demonstrating superior performance in extracting global contextual information. The vision transformer (ViT) \cite{12}, as a pioneering model that brings transformer architecture into the field of computer vision, breaks the limitations of traditional convolutional structures by employing patch partitioning and linear embedding to achieve global modeling of the entire image. Pyramid vision transformer (PVT) \cite{13} incorporates a pyramid structure to enable multi-scale feature extraction, while PVTv2 \cite{14} introduces linear attention to further reduce the computational overhead caused by the multi-scale design. LViT \cite{15} extends the ViT architecture to build cross-modal models that integrate image and text information, compensating for deficiencies in image quality through semantic guidance. Recently, numerous innovative CD methods based on CNNs and transformers have emerged \cite{50,51,52,53}, achieving impressive results in addressing various challenges. Many methods and architectures still provoke our reflection \cite{58,59,60,61,62}. However, most existing methods focus on structural features or pixel-level differences, often overlooking the complex nonlinear relationships and deep semantic information embedded in multimodal remote sensing data. To further explore the visual-linguistic understanding and interaction capabilities of large language models and to overcome the limitations of unimodal approaches, we propose to construct a text-guided multimodal model.
	
	In recent years, visual-language representation learning has achieved remarkable success, driving rapid advancements in image-text understanding within the field of computer vision. With the evolution of model architectures and pretraining strategies, tasks such as image captioning \cite{16}, text-to-image generation \cite{17}, visual question answering \cite{18}, and visual reasoning \cite{19} have flourished, becoming central topics in multimodal research. Among them, contrastive learning has been widely adopted for vision-language modeling. For instance, CLIP \cite{20} learns to embed images and texts into a shared semantic space through large-scale image-text pair training, enabling powerful zero-shot recognition capabilities. ALIGN \cite{21} builds upon this by leveraging massive web-crawled image-text data to reduce reliance on manual annotations. LLaVA \cite{22} integrates CLIP with Vicuna to construct a multimodal system capable of engaging in visual question answering. Despite their impressive performance in handling nonlinear and complex image-text data, these models still face limitations in scenarios like RSCD, which demand deeper semantic understanding and cross-temporal modeling, highlighting gaps in their generalizability and expressive power.
	
	Thus, we propose MGCR, designed for semantic-level change perception in bi-temporal remote sensing images. Specifically, we utilize the number and spatial distribution of buildings as prior cues to guide a multimodal captioning strategy based on the LLaVA model, generating structured textual descriptions from bi-temporal imagery, as illustrated in Fig. \ref{fig1}. To avoid overly verbose and complex outputs, we incorporate a regular expression-based semantic pruning strategy to extract concise, change-relevant textual content. MGCR adopts a weight-sharing siamese architecture and leverages PVT to extract visual features from the bi-temporal image pair, while semantic features are derived from the corresponding texts using CLIP’s text encoder. In the vision-language alignment stage, we introduce a graph-based conditional reconstruction module to deeply model the cross-modal semantic dependencies between image and text. This module maps multimodal features into a heterogeneous graph structure, where contextual feature interaction is achieved by constructing dependency edges between nodes. In this graph, nodes represent semantic units from different modalities, and edge weights dynamically capture cross-modal relevance, enabling both explicit alignment and implicit fusion. Finally, we integrate LViT to deeply fuse the reconstructed multimodal representations, using its hierarchical interaction mechanism to perform layered modeling and semantic refinement, thereby enhancing the model’s ability to perceive semantic changes. The main contributions of this paper are as follows: 
	
	\begin{itemize}
		\item We optimize the text generation strategy of LLaVA to produce concise and image-relevant descriptions for CD datasets, and we construct a unified multimodal CD framework MGCR for deep semantic fusion between images and text. 
		
		\item We propose a graph-based conditional reconstruction module to model semantic dependencies between visual and textual modalities, reconstructing fused features through graph attention, and further employ LViT for deep multimodal fusion to enhance semantic understanding. 
		
		\item We conduct comprehensive evaluations of MGCR on four benchmark RSCD datasets and compare it against a variety of representative methods, achieving superior performance across multiple evaluation metrics.
	\end{itemize}
	
	\begin{figure*}[!t]
		\centering
		\includegraphics[width=7in]{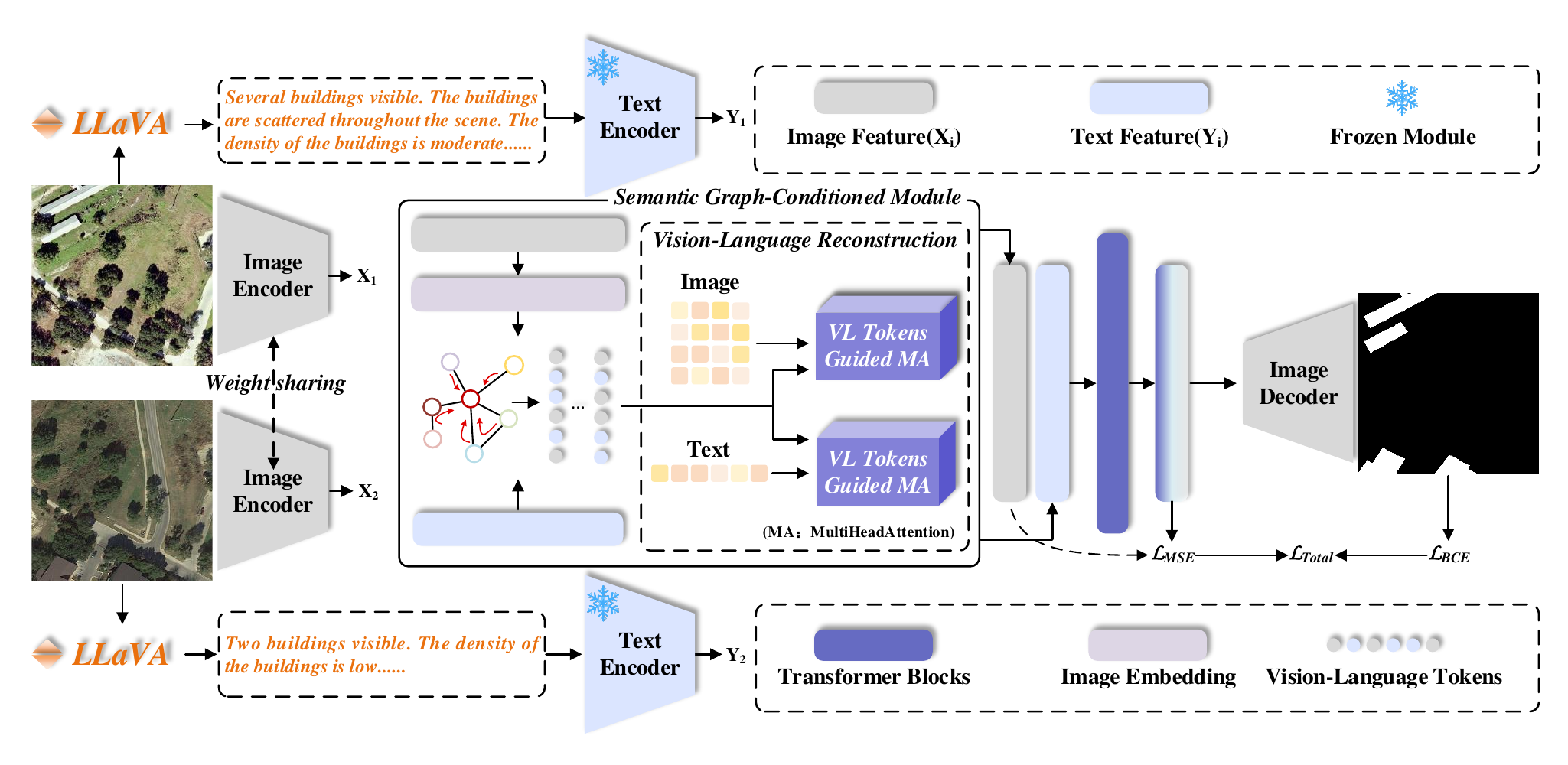}
		\caption{The overall architecture of MGCR-Net is shown. The optimized textual data generated by LLaVA is encoded using the CLIP text encoder and integrated with visual features extracted by the PVT encoder. These multimodal features are then reconstructed through SGCM and LViT modules, enabling cross-dimensional semantic information interaction.
		}
		\label{fig2}
		\vspace{-10pt}
	\end{figure*}

	\section{RELATED WORK}
	\subsection{Remote Sensing Change Detection Methods}
	With the advancement of remote sensing technologies, high-resolution imagery has further propelled the development of CD methods. From pixel-level differencing to feature-level texture extraction, traditional approaches have explored various directions in CD, yet they are often constrained by significant resource consumption. The rise of deep learning has effectively overcome these limitations. FC-EF \cite{23} is built upon the UNet architecture and implements early fusion via a fully convolutional network. FC-Siam-conc \cite{23} and FC-Siam-diff \cite{23} leverage a Siamese network with shared weights, using concatenation and differencing, respectively, to establish foundational architectures for many CD methods. IFNet \cite{24} extracts deep features via a fully convolutional structure and integrates a differential discriminator with attention mechanisms for image reconstruction. SNUNet \cite{25} introduces a densely connected Siamese network to alleviate deep feature loss via compact information transmission, and it incorporates channel attention modules for deep supervision. BIT \cite{26} employs a transformer backbone to model spatiotemporal contextual information, feeding it back into the pixel space and refining original features through a decoder. DTCDSCN \cite{27} supports both CD and semantic segmentation, incorporating dual attention modules to enhance feature representation. ChangeFormer \cite{28} synergistically combines a transformer encoder with a multi-layer perceptron decoder to effectively capture multi-scale, long-range dependencies. ICIFNet \cite{29} integrates CNNs and Transformers to achieve intra-scale cross interaction and inter-scale fusion, tapping into the full potential of hybrid models. DMINet \cite{30} unifies self-attention and cross-attention into joint attention mechanisms, enhancing global feature distribution, feature coupling, and difference detection. AERNet \cite{31} constructs an attention-guided edge refinement network that strengthens feature learning via global aggregation, attention guidance, and an adaptive loss function. SEIFNet \cite{32} introduces a spatiotemporal difference enhancement module to highlight change regions and uses an adaptive context fusion module to guide inter-layer semantic interactions. ChangeCLIP \cite{33} reconstructs original CLIP-derived features, aligns visual and textual modalities, and builds a differential compensation module to capture fine-grained semantic changes.
	
	\subsection{Development and Applications of Multimodal Large Language Models}
	In recent years, the emergence of LLM has driven artificial intelligence beyond pure language processing into multimodal understanding and generation. Typically built upon the Transformer architecture, LLM learn contextual semantic structures from massive unsupervised corpora, enabling powerful language generation and reasoning capabilities. Through instruction tuning or multi-task learning, LLM can flexibly adapt to various downstream tasks such as question answering, summarization, and dialogue, demonstrating strong generalization and contextual modeling abilities \cite{34,35,36}. In multimodal tasks that integrate vision and language, CLIP broke the limitations of traditional image classification models that rely on fixed labels by training on large-scale image-text pairs using contrastive learning. It employs separate image and text encoders to embed each modality into a shared semantic space, where semantically relevant image-text pairs are aligned with higher similarity. This cross-modal alignment mechanism enables CLIP to not only understand images but also perform open-vocabulary image retrieval and image-text matching, significantly enhancing the model's generality and scalability \cite{37,38,39}. Building upon this, LLaVA integrates CLIP’s visual perception capabilities with the language understanding and generation abilities of LLM like Vicuna, creating a powerful multimodal dialogue system \cite{40,41}. LLaVA first extracts image features via CLIP, then transforms them into a language-understandable format using a visual projection module, allowing the LLM to generate natural language outputs based on complex visual-textual inputs. This architecture not only achieves effective alignment between image and text information but also provides good extensibility, supporting multimodal tasks such as multi-turn dialogue and scene understanding. Furthermore, LLaVA's generative potential has been preliminarily validated, with researchers combining it with diffusion models for image-text generation tasks, expanding its application in the field of multimodal generation \cite{42,43}. Inspired by this, we incorporate the same principles and develop a LLaVA-based text generation optimization strategy to generate relevant textual descriptions from bitemporal remote sensing imagery.
	
	\begin{figure}[!t]
		\centering
		\includegraphics[width=3.5in]{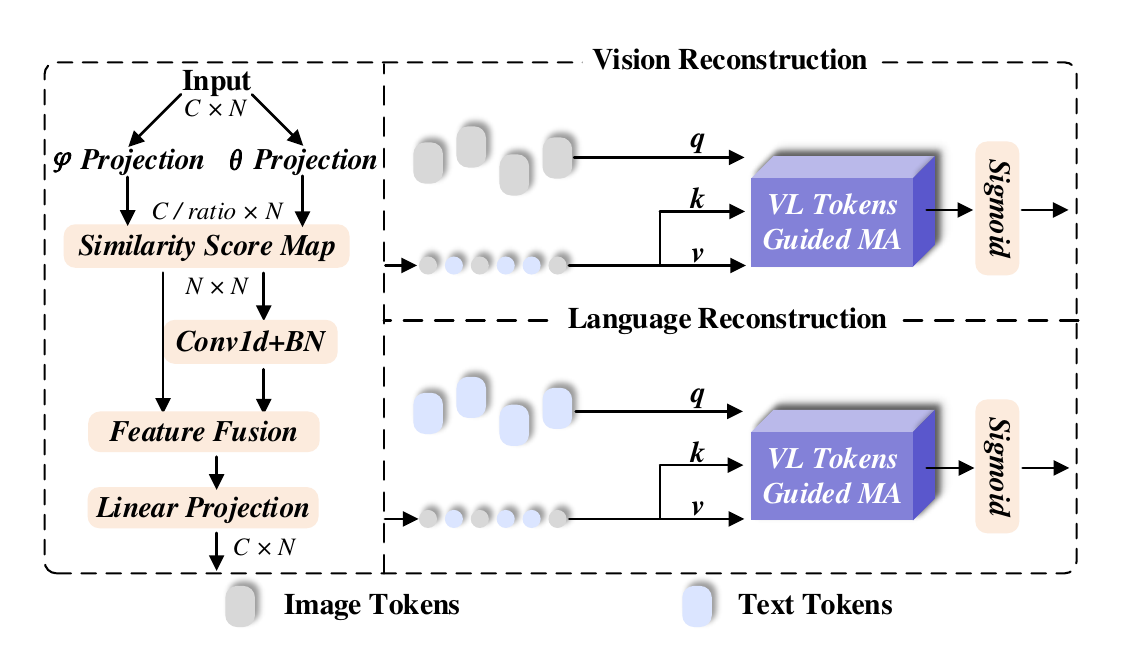}
		\caption{The demonstration illustrates the basic architecture of SGCM, where a graph attention mechanism is employed to facilitate interaction between image tokens and text tokens. Vision-Language tokens are then constructed and separately used as inputs for vision reconstruction and language reconstruction.}
		\label{fig3}
		\vspace{-10pt}
	\end{figure}

	\subsection{Attention Mechanism}
	In CD tasks, attention mechanisms enable models to focus on the differing regions of bitemporal remote sensing images, enhancing the perception of critical information. Channel attention assigns weights to feature maps across different channel dimensions to highlight key semantic channels, thereby emphasizing semantic differences between time phases. Spatial attention, on the other hand, assigns varying weights to each spatial location within the feature map, guiding the model to focus on potential change areas while suppressing irrelevant background noise. The combined use of channel and spatial attention facilitates the joint optimization of content and positional features, allowing more effective extraction of change information \cite{44,45}. To further enhance alignment and change representation between bitemporal images, cross-attention mechanisms have been widely introduced into CD tasks \cite{46,47}. Cross-attention uses features from one image as queries and features from the other as keys and values, enabling the network to establish precise semantic correspondences between the two images and better capture significant differences. Graph attention mechanisms \cite{48,49} model pixels as nodes in a graph structure, where adjacency matrices are computed to reinforce adaptive interactions between nodes, with edges representing relationships between regions. By dynamically allocating attention weights across nodes via the adjacency matrix, graph attention allows adaptive information exchange among features, thereby highlighting potential change regions and strengthening their contextual dependencies. Inspired by cross-attention, we construct the SGCM to reconstruct semantically aligned information from image-text pairs, generating VL tokens as graph condition nodes. These tokens serve as keys and values, with image and text features as queries. A multi-head attention mechanism is then employed to complete the multimodal interaction of image-text features.
	
	\section{Methodology}
	\subsection{Overall Architecture}
	In this paper, we propose an innovative method, MGCR, which leverages MLLM to focus on the CD task. As shown in Fig. \ref{fig2}, our model is composed of multiple components: LLaVA for generating textual data, PVT and CLIP as image-text encoders, SGCM for conditional reconstruction and alignment between modalities, and a LViT-based structure for deep fusion of visual and textual features. First, we utilize LLaVA’s image-text understanding and powerful user interaction capabilities to simulate a dialogue process in which it responds to prompts about image content. From the extensive textual descriptions generated, we apply regular expression-based pruning to retain only text related to building count and spatial distribution, thereby constructing multimodal input data specifically tailored for building CD. Next, we employ PVT and CLIP models as encoders to extract features, using PVT’s pyramid architecture to derive multi-scale features from bitemporal images, while feeding the generated text into CLIP’s text encoder to obtain textual representations. To effectively combine image and text features, we design SGCM based on graph attention, which models the correspondence between image regions and text tokens. This module treats salient image regions as graph nodes and uses textual guidance to construct adjacency relations, enabling structured interactions and feature reconstruction under semantic alignment, thus overcoming the limitations of unimodal CD by leveraging multidimensional information. Finally, in the cross-modal feature integration stage, we incorporate the LViT structure to perform deep fusion of image and text features. Through hierarchical self-attention, this structure dynamically learns semantic associations between modalities, enhancing the model’s ability to discern building edges and morphological differences, and providing fine-grained semantic information.

	\begin{figure*}[h]
		\centering
		\includegraphics[width=7in]{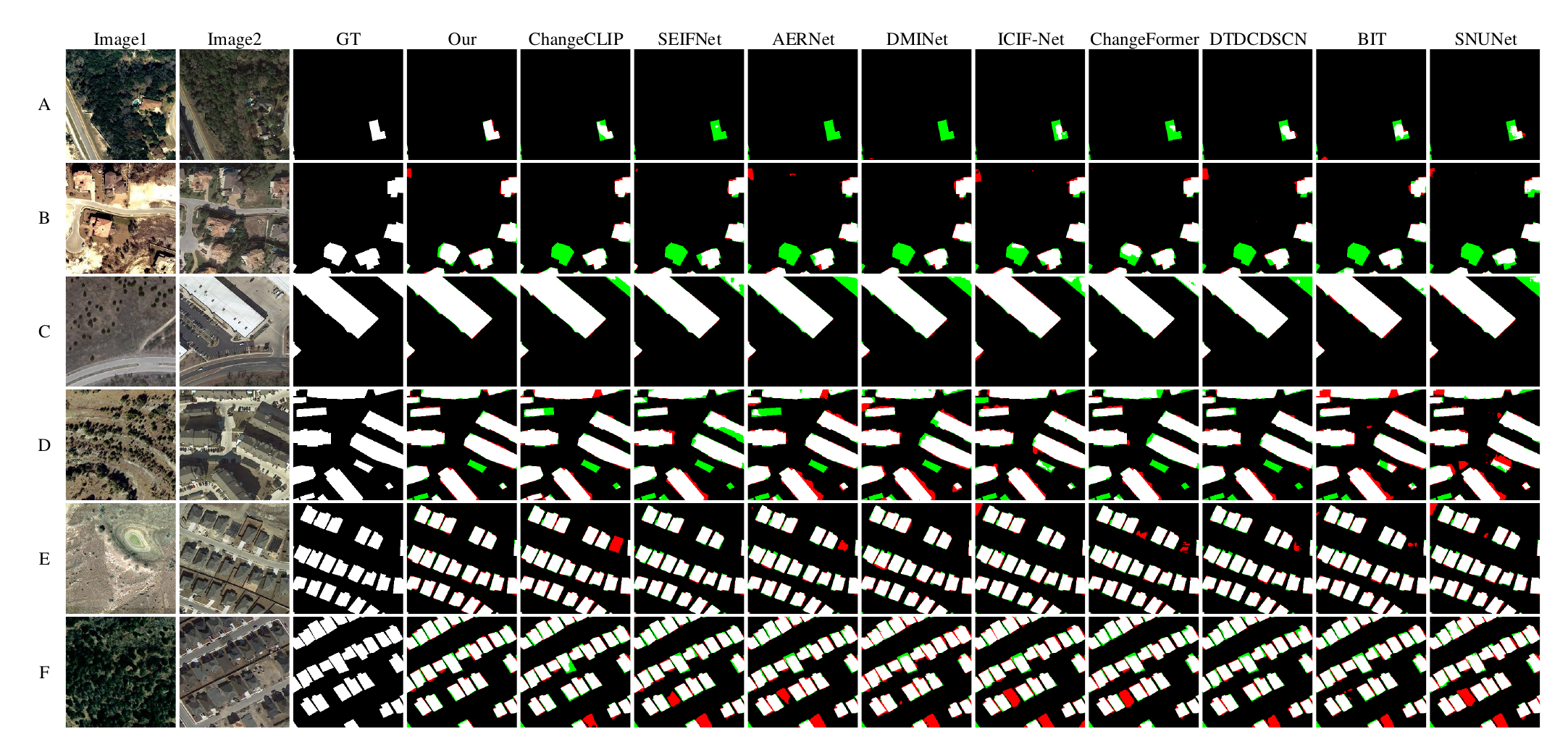}
		\caption{The comparison of the prediction results of MGCR-Net with mainstream methods on the LEVIR-CD dataset. In this figure, the green areas represent the predicted parts that are missing compared to the GT, while the red areas represent the redundant predicted parts compared to GT.}
		\label{fig4}
		\vspace{-10pt}
	\end{figure*}

	\subsection{Text Generation Based on Multimodal Large Models}
	In this paper, we leverage the powerful image-text alignment and language modeling capabilities of LLaVA as a multimodal text generator. After task-specific optimization of LLaVA for building CD, we input the raw images, and LLaVA generates descriptive text containing structured semantic information, including the number of buildings, their spatial distribution, and density. Additionally, it provides judgments and descriptions of vehicles, roads, grass, trees, and other elements. The simplified mathematical formulation is as follows:
	
	\begin{equation}\label{eq:1}
	\begin{gathered}
	V=\mathcal{E}_v\left( X \right),
	\end{gathered}
	\end{equation}
	\begin{equation}\label{eq:2}
	\begin{gathered}
	V^{'}=W_{pV}+b_{p},
	\end{gathered}
	\end{equation}
	\begin{equation}\label{eq:3}
	\begin{gathered}
	\hat{y}_{t}=\text{MLLM}\left( Concat\left( V^{'},T_{prompt} \right) \right),
	\end{gathered}
	\end{equation}
	here, X represents the raw image input, $\mathcal{E}_{v}$ denotes the visual encoder of CLIP, and $\mathcal{E}_{v}$ extracts the visual features of the image to obtain V. $W_{p}$ is a trainable linear projection matrix, and $b_{p}$ is a trainable bias term. By linearly mapping the visual features and aligning them with the feature space of the language model, we obtain $V^{'}$, which is then integrated with $T_{prompt}$ and passed through the MLLM to generate the predicted output $\hat{y}_{t}$.
	
	\begin{table}[h]
		\caption{Indicator results for the LEVIR-CD dataset. Red color represents the best results and blue color represents the second best results (\%).\label{tab1}}
		\centering
		\renewcommand\arraystretch{1.2}
		\resizebox{0.49\textwidth}{!}
		{
			\begin{tabular}{ccccc}
				\hline
				Methods        & F1 & IoU & Precision & Recall \bigstrut[t] \\
				\hline
				SNUNet\cite{25}         & 88.16 & 78.83 & 89.18 & 87.17 \bigstrut[t] \\
				BIT\cite{26}            & 89.31 & 80.68 & 89.24 & 89.37 \bigstrut[t] \\
				DTCDSCN\cite{27}        & 87.67 & 78.05 & 88.53 & 86.83 \bigstrut[t] \\
				ChangeFormer\cite{28}   & 90.40 & 82.48 & 92.05 & 88.80 \bigstrut[t] \\
				ICIF-Net\cite{29}       & 91.18 & 83.85 & 91.13 & 90.57 \bigstrut[t] \\
				DMINet\cite{30}         & 90.71 & 82.99 & 92.52 & 89.95 \bigstrut[t] \\
				AERNet\cite{31}         & 90.78 & 83.11 & 89.97 & \textcolor{red}{91.59} \bigstrut[t] \\
				SEIFNet\cite{32}        & 90.86 & 83.25 & \textcolor{red}{94.29} & 87.67 \bigstrut[t] \\
				ChangeCLIP\cite{33}	   & \textcolor{blue}{91.89} & \textcolor{blue}{85.01} & \textcolor{blue}{92.93} & 90.89 \bigstrut[t] \\
				Ours           & \textcolor{red}{92.07} & \textcolor{red}{85.30} & 92.52 & \textcolor{blue}{91.22} \bigstrut[t] \\
				\hline
			\end{tabular}
		}
		\vspace{-10pt}
	\end{table}

	\subsection{The Feature Encoder}
	PVT is a Transformer architecture designed for computer vision tasks, with its core advantage lying in the introduction of a pyramid structure that endows the Transformer with multi-scale feature extraction capabilities similar to CNNs. PVT divides the input image into smaller patches and progressively reduces spatial resolution while enhancing semantic features layer by layer, achieving hierarchical feature representation. Each scale consists of multiple stages, where a spatial-reduction attention mechanism is employed to effectively reduce the computational cost of self-attention, enabling PVT to efficiently handle high-resolution images while preserving the long-range modeling capabilities of Transformers. Ultimately, PVT outputs multi-resolution pyramid features, offering rich multi-scale semantic representations, and its mathematical formulation is as follows:
	\begin{equation}\label{eq:4}
	\begin{gathered}
	X_{ij}=E_i\left( X_i \right),
	\end{gathered}
	\end{equation}
	here, E denotes the PVT encoder, i represents the siamese network processing the bi-temporal images, and j refers to the different scales obtained after feature extraction by the PVT encoder.
	
	\begin{figure*}[t]
		\centering
		\includegraphics[width=7in]{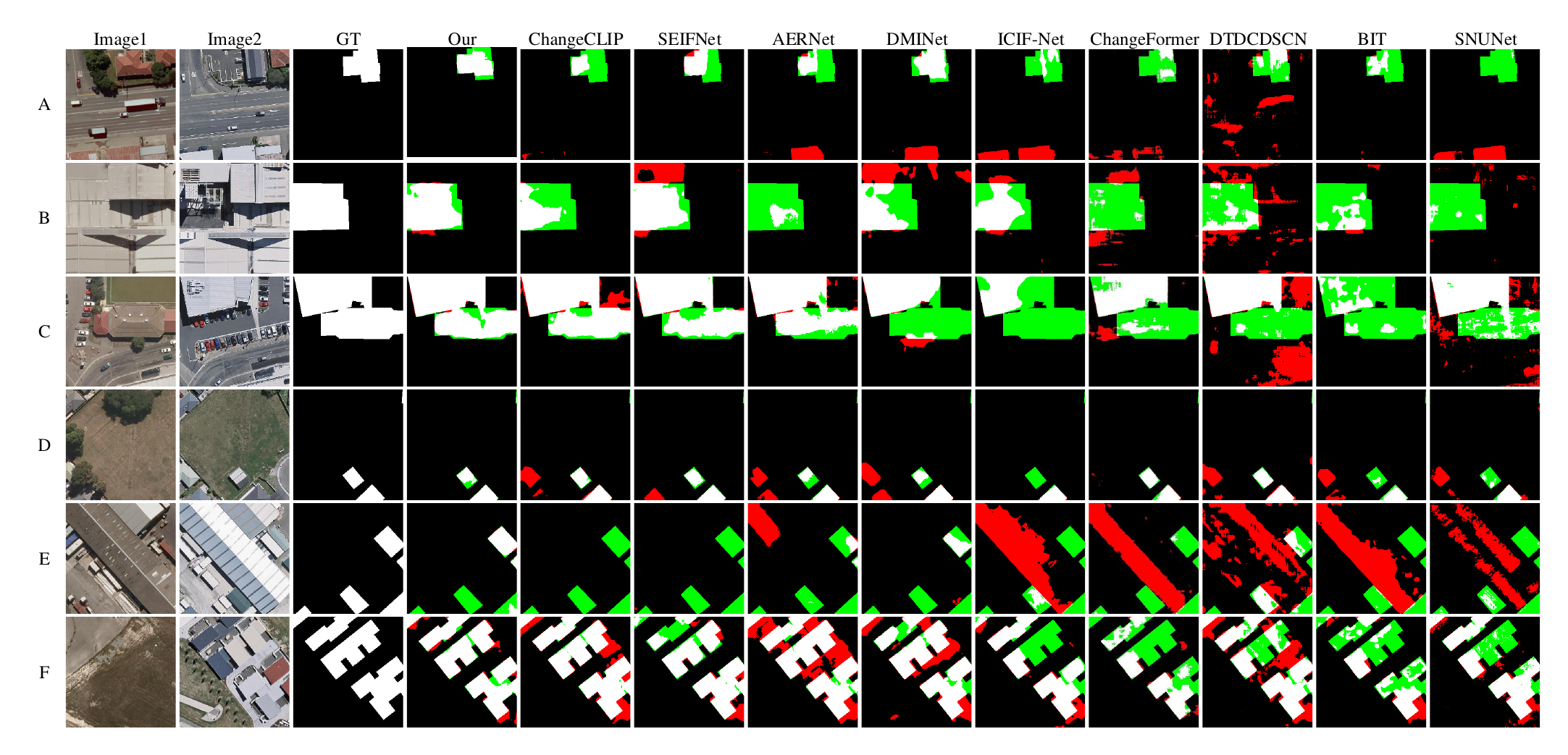}
		\caption{The comparison of the prediction results of MGCR-Net with mainstream methods on the WHU-CD dataset. In this figure, the green areas represent the predicted parts that are missing compared to the GT, while the red areas represent the redundant predicted parts compared to GT.}
		\label{fig5}
		\vspace{-10pt}
	\end{figure*}

	CLIP is a general vision-language pretraining architecture based on cross-modal contrastive learning. Its core idea is to encode images and natural language texts into a shared semantic embedding space by maximizing the similarity between matching image-text pairs and minimizing the similarity between mismatched pairs, thereby achieving semantic alignment across modalities. Unlike traditional classification models that rely on manually annotated labels, CLIP is trained on large-scale image-text pairs without predefined category labels, allowing it to automatically capture deep semantic relationships between visual and linguistic information and exhibit strong open-vocabulary understanding capabilities. In this paper, we fully leverage CLIP’s strengths in cross-modal modeling, specifically adopting its text encoder to transform natural language prompts into semantic vectors that can be aligned with visual features. This encoder not only comprehends word-level semantics but also captures sentence-level contextual structures, providing fine-grained modulation signals for image semantics. Compared to conventional visual encoders, CLIP’s text embeddings carry rich linguistic priors learned during training, enabling the model to understand targets and relationships in images based on flexible natural language descriptions rather than being restricted to a fixed label set. This offers stronger semantic support for subsequent multimodal fusion and CD tasks. The corresponding mathematical formulation is as follows:
	\begin{equation}\label{eq:5}
	\begin{gathered}
	T_i=\mathcal{E}_{ti}\left( \hat{y}_{ti} \right),
	\end{gathered}
	\end{equation}
	where $\mathcal{E}_{ti}$ denotes the text encoder of CLIP, and $T_i$ represents the text features obtained by encoding the textual descriptions of the bi-temporal images using the text encoder.
	
	\begin{table}[h]
		\caption{Indicator results for the WHU-CD dataset. Red color represents the best results and blue color represents the second best results (\%).\label{tab2}}
		\centering
		\renewcommand\arraystretch{1.2}
		\resizebox{0.49\textwidth}{!}
		{
			\begin{tabular}{ccccc}
				\hline
				Methods         & F1 & IoU & Precision & Recall \bigstrut[t] \\
				\hline
				SNUNet\cite{25}         & 88.34 & 79.11 & 91.34 & 85.53 \bigstrut[t] \\
				BIT\cite{26}            & 87.47 & 77.73 & 88.71 & 86.27 \bigstrut[t] \\
				DTCDSCN\cite{27}        & 90.48 & 82.62 & 91.84 & 89.16 \bigstrut[t] \\
				ChangeFormer\cite{28}   & 86.88 & 76.81 & 88.50 & 85.33 \bigstrut[t] \\
				ICIF-Net\cite{29}       & 90.77 & 83.09 & 92.93 & 88.70 \bigstrut[t] \\
				DMINet\cite{30}         & 91.49 & 84.31 & 92.65 & 90.35 \bigstrut[t] \\
				AERNet\cite{31}         & 92.18 & 85.49 & 92.47 & 91.89 \bigstrut[t] \\
				SEIFNet\cite{32}        & 93.29 & 87.43 & 93.99 & 92.61 \bigstrut[t] \\
				ChangeCLIP\cite{33}     & \textcolor{blue}{94.40} & \textcolor{blue}{89.40} & \textcolor{blue}{95.18} & \textcolor{red}{93.64} \bigstrut[t] \\
				Ours           & \textcolor{red}{94.91} & \textcolor{red}{90.32} & \textcolor{red}{96.42} & \textcolor{blue}{93.45} \bigstrut[t] \\
				\hline
			\end{tabular}
		}
	\end{table}
	
	\begin{figure*}[h]
		\centering
		\includegraphics[width=7in]{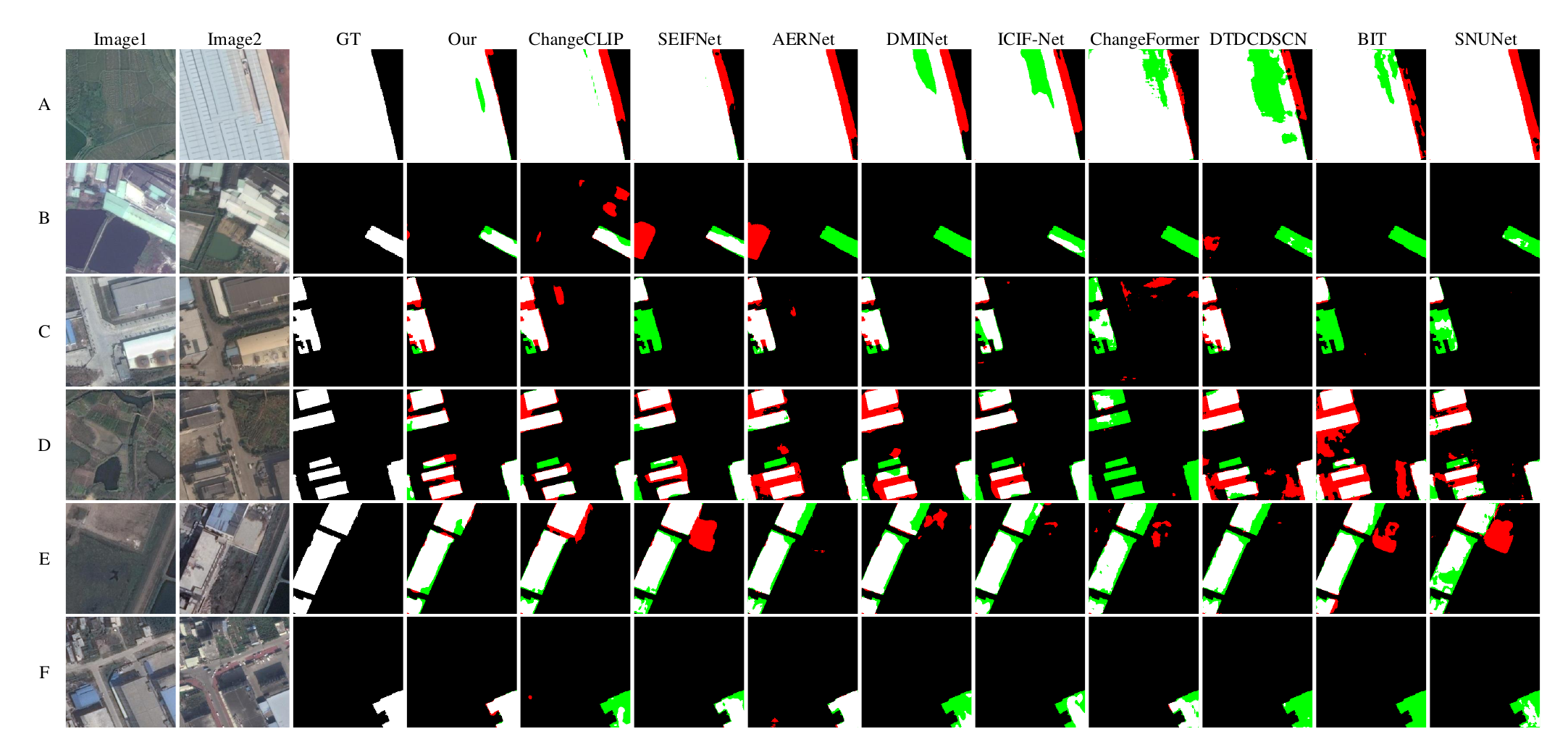}
		\caption{The comparison of the prediction results of MGCR-Net with mainstream methods on the GZ-CD dataset. In this figure, the green areas represent the predicted parts that are missing compared to the GT, while the red areas represent the redundant predicted parts compared to GT.}
		\label{fig6}
		\vspace{-10pt}
	\end{figure*}

	\subsection{Semantic Graph-Conditioned Module}
	In multimodal CD tasks, the effective integration of visual and textual features is crucial. Misalignment between visual and textual features can negatively impact the model’s performance. Cross-attention-based fusion is a commonly used approach to align visual and textual features in multimodal tasks. The SGCM consists of three components, as shown in the Fig. \ref{fig3},  first, graph attention is applied to achieve an initial fusion of visual and textual features, resulting in VL tokens. Then, two graph-conditioned reconstruction modules are constructed—one for text-to-vision and the other for vision-to-text. These modules further enhance the semantic association between vision and language by automatically learning to reconstruct multimodal features during training.
	\begin{equation}\label{eq:6}
	\begin{gathered}
	E_{X_{i3}}=\text{Embedding}\left( X_{i3} \right) +E_{pos},
	\end{gathered}
	\end{equation}	
	
	\begin{equation}\label{eq:7}
	\begin{gathered}
	VL=\text{Concat}\left( E_{X_{i3}},T_i \right),
	\end{gathered}
	\end{equation}	
	we adopt a single-layer interaction feature alignment strategy, where the fusion and alignment of visual and textual features are built upon the deep pyramid structure with global attribute features. Therefore, we select $X_{i3}$ as the image input and apply an embedding operation along with the addition of positional encoding $E_{pos}$ to achieve an initial structural alignment between the image and text. Subsequently, the aligned features are concatenated to obtain VL.
	
	\begin{table}[h]
		\caption{Indicator results for the GZ-CD dataset. Red color represents the best results and blue color represents the second best results (\%).\label{tab3}}
		\centering
		\renewcommand\arraystretch{1.2}
		\resizebox{0.49\textwidth}{!}
		{
			\begin{tabular}{ccccc}
				\hline
				Methods         & F1 & IoU & Precision & Recall \bigstrut[t] \\
				\hline
				SNUNet\cite{25}         & 84.25 & 72.79 & 84.25 & 81.82 \bigstrut[t] \\
				BIT\cite{26}            & 80.23 & 66.99 & 82.40 & 78.18 \bigstrut[t] \\
				DTCDSCN\cite{27}        & 83.00 & 70.93 & 88.19 & 78.38 \bigstrut[t] \\
				ChangeFormer\cite{28}   & 73.66 & 58.30 & 84.59 & 65.23 \bigstrut[t] \\
				ICIF-Net\cite{29}       & 85.09 & 74.05 & 89.90 & 80.76 \bigstrut[t] \\
				DMINet\cite{30}         & 81.98 & 69.46 & 87.92 & 76.79 \bigstrut[t] \\
				AERNet\cite{31}         & 84.42 & 73.03 & 88.06 & 81.07 \bigstrut[t] \\
				SEIFNet\cite{32}        & 87.48 & 77.75 & 89.64 & \textcolor{blue}{85.43} \bigstrut[t] \\
				ChangeCLIP\cite{33}     & \textcolor{blue}{88.38} & \textcolor{blue}{79.18} & \textcolor{blue}{91.86} & 85.15 \bigstrut[t] \\
				Ours           & \textcolor{red}{89.55} & \textcolor{red}{81.07} & \textcolor{red}{92.07} & \textcolor{red}{87.16} \bigstrut[t] \\
				\hline
			\end{tabular}
		}
	\end{table}

	\begin{equation}\label{eq:8}
	\begin{gathered}
	\varPhi =\text{LayerNorm}\left( \phi \left( VL \right) \right),
	\end{gathered}
	\end{equation}
	\begin{equation}\label{eq:9}
	\begin{gathered}
	\varTheta =\text{LayerNorm}\left( \theta \left( VL \right) \right),
	\end{gathered}
	\end{equation}
	\begin{equation}\label{eq:10}
	\begin{gathered}
	z=\varPhi \cdot \varTheta ^T,
	\end{gathered}
	\end{equation}
	\begin{equation}\label{eq:11}
	\begin{gathered}
	\mathcal{Z}=\text{BN}\left( \text{Conv1D}\left( z^T \right) \right) ^T+z,
	\end{gathered}
	\end{equation}

	first, the fused representation VL undergoes two linear transformations to obtain $\varPhi$ and $\varTheta$, which represent the relationships between node pairs. Then, an adjacency matrix $z$ is constructed along the sequence dimension to model the similarity score map between visual and textual features. By integrating the graph concept with convolution operations, the overall graph of key vision-language nodes $\mathcal{Z}$ is constructed.
	\begin{equation}\label{eq:12}
	\begin{gathered}
	F=\text{Concat}\left( z,\mathcal{Z} \right),
	\end{gathered}
	\end{equation}
	\begin{equation}\label{eq:13}
	\begin{gathered}
	F^{'}=\text{BN}\left( \text{Conv1D}\left( F^{T} \right) \right) ^T,
	\end{gathered}
	\end{equation}
	\begin{equation}\label{eq:14}
	\begin{gathered}
	VL\ Tokens_{}=\text{ReLU}\left( VL+\text{Proj}\left( F^{'} \right) \right),
	\end{gathered}
	\end{equation}
	Finally, the VL tokens are projected into the feature space.
	\begin{equation}\label{eq:15}
	\begin{gathered}
	K=V=VL\ Tokens,
	\end{gathered}
	\end{equation}
	\begin{equation}\label{eq:16}
	\begin{gathered}
	VR=\text{softmax} \left( \frac{Q_{img}K^{T}}{\sqrt{d}} \right) V,
	\end{gathered}
	\end{equation}
	\begin{equation}\label{eq:17}
	\begin{gathered}
	LR=\text{softmax} \left( \frac{Q_{text}K^{T}}{\sqrt{d}} \right) V,
	\end{gathered}
	\end{equation}
	where VR denotes vision reconstruction and LR denotes language reconstruction. We employ a cross-attention mechanism to build semantic associations between the two modalities, using VL Tokens as the keys and values , while using the image and text features respectively as the queries. In this way, the image and text can autonomously learn contextual information from the VL Tokens, the image representation is enhanced through textual guidance, while the textual representation is adjusted using visual information, enabling collaborative semantic modeling between vision and language.
	
		\begin{figure*}[!t]
		\centering
		\includegraphics[width=7in]{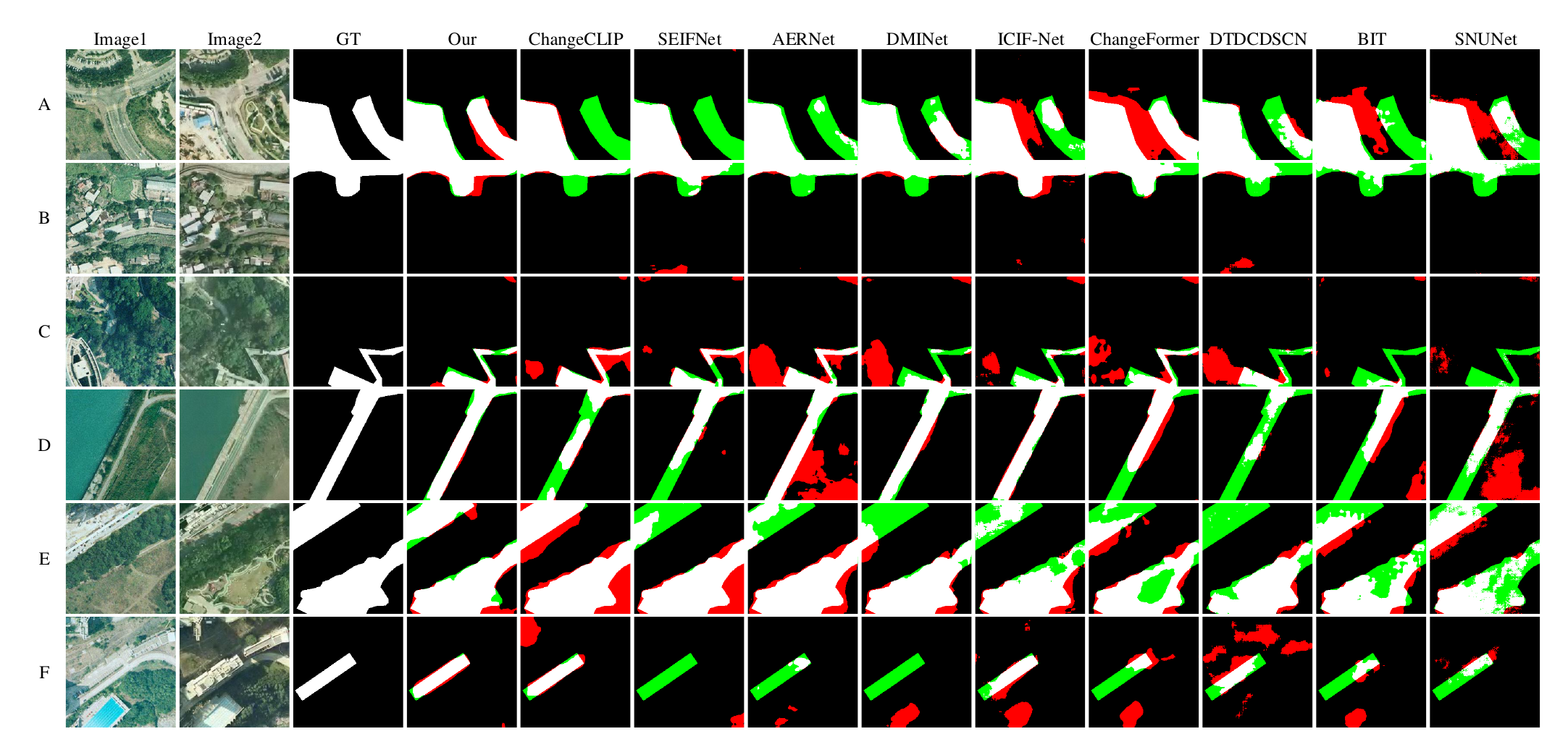}
		\caption{The comparison of the prediction results of MGCR-Net with mainstream methods on the SYSU-CD dataset. In this figure, the green areas represent the predicted parts that are missing compared to the GT, while the red areas represent the redundant predicted parts compared to GT.
		}
		\label{fig7}
		\vspace{-10pt}
	\end{figure*}

	\begin{equation}\label{eq:18}
	\begin{gathered}
	X_{i3}=X_{i3}+\text{sigmoid}\left( VR \right),
	\end{gathered}
	\end{equation}
	\begin{equation}\label{eq:19}
	\begin{gathered}
	T_{i}=T_{i}+\text{sigmoid}\left( LR \right),
	\end{gathered}
	\end{equation}
	\begin{equation}\label{eq:20}
	\begin{gathered}
	VL=\text{LViT}\left( X_{i3},T_{i} \right),
	\end{gathered}
	\end{equation}
	we leverage LViT to further model the vision-language features, aiming to deeply fuse the cross-modal semantic relationships between image and text. Specifically, after the initial alignment and reconstruction of vision-language features, we feed them into a multi-layer transformer encoder module to provide a richer and more semantically consistent representation for subsequent CD.
	
	\begin{table}[h]
		\caption{Indicator results for the SYSU-CD dataset. Red color represents the best results and blue color represents the second best results (\%).\label{tab4}}
		\centering
		\renewcommand\arraystretch{1.2}
		\resizebox{0.49\textwidth}{!}
		{
			\begin{tabular}{ccccc}
				\hline
				Methods          & F1 & IoU & Precision & Recall \bigstrut[t] \\
				\hline
				SNUNet\cite{25}         & 73.14 & 57.66 & 74.09 & 72.21 \bigstrut[t] \\
				BIT\cite{26}            & 74.32 & 59.13 & 81.75 & 68.12 \bigstrut[t] \\
				DTCDSCN\cite{27}        & 77.22 & 62.89 & 80.42 & 74.26 \bigstrut[t] \\
				ChangeFormer\cite{28}   & 74.41 & 59.24 & 77.99 & 71.14 \bigstrut[t] \\
				ICIF-Net\cite{29}       & 76.08 & 61.40 & 79.16 & 73.24 \bigstrut[t] \\
				DMINet\cite{30}         & 80.06 & 66.75 & 84.93 & 75.72 \bigstrut[t] \\
				AERNet\cite{31}         & 80.26 & 67.02 & 82.50 & 78.13 \bigstrut[t] \\
				SEIFNet\cite{32}        & 81.46 & 68.72 & 83.59 & \textcolor{blue}{79.44} \bigstrut[t] \\
				ChangeCLIP\cite{33}     & \textcolor{blue}{82.61} & \textcolor{blue}{70.38} & \textcolor{red}{86.64} & 78.94 \bigstrut[t] \\
				Ours           & \textcolor{red}{82.91} & \textcolor{red}{70.80} & \textcolor{blue}{86.52} & \textcolor{red}{79.58} \bigstrut[t] \\
				\hline
			\end{tabular}
		}
	\end{table}
	
	\section{Experimental results and analysis}
	Our optimized text generation data is specifically designed for building CD. To further verify the effectiveness and generalization ability of MGCR in broader multimodal CD tasks, we conduct experiments on publicly available datasets not limited to building CD, including LEVIR-CD \cite{54}, WHU-CD \cite{55}, GZ-CD \cite{56}, and additionally SYSU-CD \cite{57}. We perform comparative experiments with several mainstream methods as well as ablation studies on MGCR to comprehensively evaluate the performance of the proposed model.
	\subsection{Datasets}
	\subsubsection{LEVIR-CD Dataset}
	The LEVIR-CD dataset is specifically designed for building CD across diverse environments, including urban and suburban scenes. It consists of 637 image pairs, each with a resolution of $1024\times 1024$ pixels and annotated at the pixel level. We randomly crop the original images into $256\times 256$ patches and split them into training, validation, and test sets with a ratio of 7:2:1. This dataset features multi-scale changes, diverse surface cover types, and complex backgrounds, making it suitable for evaluating the generalization and robustness of the model.

	\subsubsection{WHU-CD Dataset}
	The WHU-CD dataset focuses on building CD in the context of urban expansion. It includes various land cover types such as buildings, roads, and green areas, with fine-grained pixel-level annotations. The images are cropped into $256\times 256$ patches, yielding a total of 7620 image pairs, which are split into training, validation, and test sets at a ratio of 8:1:1. This dataset is suitable for assessing the model’s adaptability to different urban textures and surface change patterns.
	
	\subsubsection{GZ-CD Dataset}
	The GZ-CD dataset targets building CD in densely populated urban areas. It contains tasks involving both large-scale structures and small scattered buildings, while also presenting complex cases such as building shadows and overlapping regions. The images are cropped into $256\times 256$ patches, with 2834 pairs for training, 400 for validation, and 325 for testing. This dataset is valuable for evaluating the model’s ability to detect subtle changes in complex urban scenarios.
	
	\subsubsection{SYSU-CD Dataset}
	Released by Sun Yat-sen University, the SYSU-CD dataset is a high-resolution RSCD dataset featuring well-aligned image pairs and accurate labels. It includes typical urban building changes as well as other types of changes such as vegetation shifts and road expansion. Each image pair is sized $256\times 256$, and the dataset contains a total of 20,000 pairs, split into training, validation, and test sets in an 8:1:1 ratio.
	
	\subsection{Implementation Environment}
	The experiments were conducted on an Ubuntu 18.04 system for code execution and debugging. Model development and training were performed using the PyTorch deep learning framework, accelerated by an NVIDIA TITAN RTX 24GB GPU. During the optimization of LLaVA’s strategy, the following parameters were set, temperature=0.1, top\_p=0.3, num\_beams=5, and max\_new\_tokens=256. We provided prompts focused on building-related content to generate rich multimodal text data. For other types of CD tasks, such as SYSU-CD, we adopted a text data generation method similar to ChangeCLIP, using the probabilities of multiple background categories as text inputs. AdamW was selected as the optimizer with a learning rate of 5e-4 and a weight decay of 0.01. A cosine annealing schedule was applied for learning rate adjustment. The model training was limited to a maximum of 300 epochs. For guiding model optimization, we combined binary cross-entropy loss and mean squared error loss. The complete loss structure can be expressed as:
	\begin{equation}\label{eq:21}
	\begin{aligned}
	L_{BCE}=-\frac{1}{N}\sum_{i=1}^{N}{\left[ y_{i}\log \left( \hat{y}_{i} \right) +\left( 1-y_{i} \right) \log \left( 1-\hat{y}_{i} \right) \right]},
	\end{aligned}
	\end{equation}
	\begin{equation}\label{eq:22}
	\begin{aligned}
	L_{MSE}=\frac{1}{N}\sum_{i=1}^{N}{\left( y_{i}-\hat{y}_{i} \right) ^{2}},
	\end{aligned}
	\end{equation}
	\begin{equation}\label{eq:23}
	\begin{aligned}
	L=\lambda _{1}L_{BCE}+\left( \lambda _{2}+\lambda _{3} \right) L_{MSE},
	\end{aligned}
	\end{equation}
	here, $\lambda_{1}$, $\lambda_{2}$, and $\lambda_{3}$ are weighting hyperparameters. $L_{BCE}$ supervises the MGCR-generated predicted maps against GT, while $L_{MSE}$ measures the similarity between the features of the bi-temporal images and the fused multimodal features, enhancing the consistency of the multimodal feature representations.
	
	\subsection{Evaluation Metrics}
	To evaluate the performance of our model on the CD task and compare it with other methods, we employ four commonly used metrics, F1, IoU, precision, and recall. The F1, being the harmonic mean of precision and recall, is especially suitable for imbalanced class scenarios and provides a more comprehensive performance assessment. IoU measures the spatial overlap between the predicted change regions and the ground truth change labels, reflecting the detection accuracy of the model. Precision indicates the proportion of correctly predicted positive samples among all predicted positives, while recall measures the model’s ability to identify all true positive samples. The specific mathematical formulas for these metrics are provided below:
	\begin{equation}\label{eq:24}
	\begin{gathered}
	F1=\frac{2TP}{2TP+FP+FN},
	\end{gathered}
	\end{equation}
	\begin{equation}\label{eq:25}
	\begin{gathered}
	IoU=\frac{TP}{TP+FP+FN},
	\end{gathered}
	\end{equation}
	\begin{equation}\label{eq:26}
	\begin{gathered}
	Precision=\frac{TP}{TP+FP},
	\end{gathered}
	\end{equation}
	\begin{equation}\label{eq:27}
	\begin{gathered}
	Recall=\frac{TP}{TP+FN},
	\end{gathered}
	\end{equation}
	where TP represents true positives, FP represents false positives, FN represents false negatives, and TN represents true negatives.
	
	\subsection{Comparative Experiment}
	In this section, we evaluate the effectiveness of MGCR on four CD datasets: LEVIR-CD, WHU-CD, GZ-CD, and SYSU-CD. Throughout the comparison with other mainstream methods, MGCR consistently achieves the best performance both in terms of visualized prediction maps and commonly used quantitative metrics.
	
	As shown in the Fig. \ref{fig4}, we selected representative samples A–F from the LEVIR-CD dataset based on characteristics such as building quantity and density, choosing different perspectives for detailed analysis. Samples A and B have backgrounds dominated by roads and trees. In sample A, buildings surrounded by trees with colors similar to the foliage were difficult for many mainstream methods to distinguish as change targets. ChangeCLIP also showed imprecise detection of difference regions due to blurred boundaries between buildings and trees. In sample B, the building in the lower-left corner blends into the surrounding background, causing almost all compared methods to fail in locating the building, which led to failed CD. However, our MGCR method not only accurately located the building at this position but also predicted the precise shape of the difference region nearly completely. In sample C, MGCR’s prediction was closest to the GT, with only minor errors along the edges of the predicted target area, while other methods produced blurred localizations and considerable fragmentation in detecting the difference region in the upper right. For samples E and F, there were parts of detected difference regions that differed significantly from actual change areas; these false positives often resembled true difference areas but were non-building regions. Methods like ChangeCLIP, SEIFNet, and AERNet exhibited redundant predictions in these areas. As shown in the Table. \ref{tab1}, we conducted comparisons across multiple mainstream methods using four common evaluation metrics corresponding to the visualized results. Our method outperformed the second-best, ChangeCLIP, with improvements of 0.18\% in F1 and 0.29\% in IoU.
	
	\begin{figure*}[!t]
		\centering
		\includegraphics[width=7in]{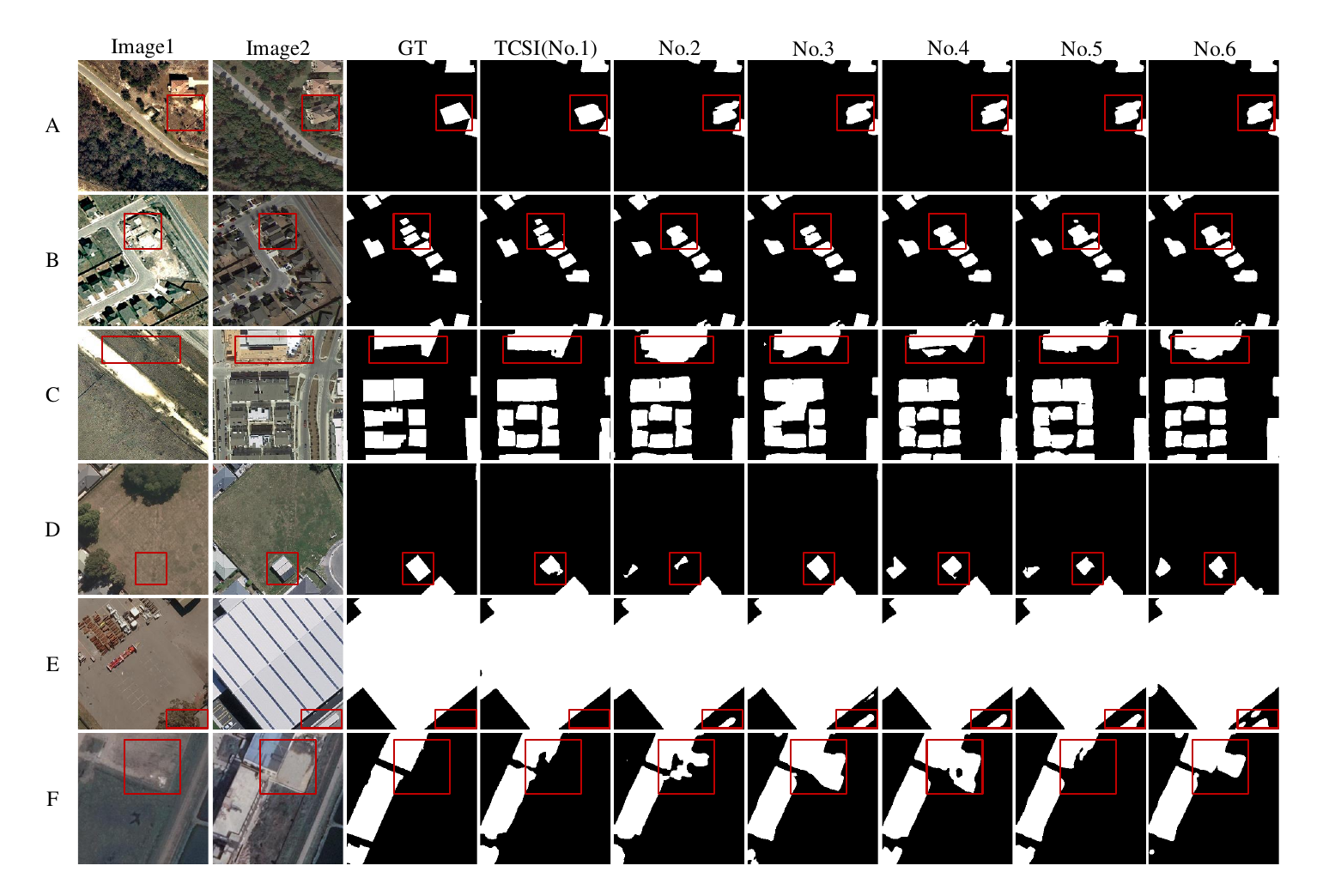}
		\caption{Ablation experiments of MGCR on four datasets, with key focus areas in the samples highlighted using red boxes.}
		\label{fig8}
		\vspace{-10pt}
	\end{figure*}

	\begin{table*}[!ht]
		\centering
		\setlength{\tabcolsep}{3pt}
		\caption{\textrm{Overview of ablation results for MGCR-Net and metric outcomes on four datasets (\%).}}
		\begin{tabular}{ccccccccccccc}
			\hline
			\multirow{2}{*}{No.} & \multirow{2}{*}{SGCM-L} & \multirow{2}{*}{SGCM-V} & \multirow{2}{*}{LViT} & \multicolumn{2}{c}{\textbf{LEVIR-CD}} & \multicolumn{2}{c}{\textbf{WHU-CD}} & \multicolumn{2}{c}{\textbf{GZ-CD}}  & \multicolumn{2}{c}{\textbf{SYSU-CD}} \bigstrut[t] \\
			\cline{5-12}
			& & & & \textbf{F1} & \textbf{IoU} & \textbf{F1} & \textbf{IoU} & \textbf{F1} & \textbf{IoU} & \textbf{F1} & \textbf{IoU} \bigstrut[t] \\
			\hline
			\textbf{1}   & \checkmark & \checkmark & \checkmark 
			& \textbf{92.07}/91.91 & \textbf{85.30}/85.03 & \textbf{94.91}/94.65 & \textbf{90.32}/89.85 & \textbf{89.55}/89.35 & \textbf{81.07}/80.74 & \textbf{82.91}/82.76 & \textbf{70.80}/70.60 \bigstrut[t] \\
			\textbf{2}   & \checkmark & $\times$ & \checkmark 
			& 91.77/91.75 & 84.79/84.76 & 94.60/94.43 & 89.76/89.45 & 88.83/88.70 & 79.91/79.69 & 82.50/82.33 & 70.21/69.96 \bigstrut[t] \\
			\textbf{3}   & $\times$ & \checkmark & \checkmark
			& 91.67/91.73 & 84.62/84.72 & 94.53/94.46 & 89.62/89.50 & 89.04/88.82 & 80.25/79.88 & 82.55/82.43 & 70.29/70.11 \bigstrut[t] \\
			\textbf{4}   & $\times$ & $\times$ & \checkmark 
			& 91.58/91.47 & 84.47/84.28 & 94.38/94.24 & 89.36/89.10 & 88.38/88.35 & 79.18/79.13 & 82.19/82.02 & 69.77/69.52 \bigstrut[t] \\
			\textbf{5}   & \checkmark & \checkmark & $\times$
			& 91.58/- & 84.47/- & 94.28/- & 89.18/- & 88.72/- & 79.72/- & 82.27/- & 69.88/- \bigstrut[t] \\
			\textbf{6}   & $\times$ & $\times$ & $\times$
			& 91.25/- & 83.90/- & 93.87/- & 88.44/- & 88.06/- & 78.67/- & 81.76/- & 69.15/- \bigstrut[t] \\
			\hline
		\end{tabular}
		\label{tab5}
	\end{table*}
	
	As shown in the Fig. \ref{fig5}, we selected representative samples A–F from the WHU-CD dataset based on characteristics such as building quantity and density, considering diverse perspectives. We provide detailed analysis of several samples. In sample A, the bi-temporal images depict structural renovations at corresponding locations, resulting in a relatively complex form of building change. In contrast, our method, along with DMINet, delivered relatively good performance. Sample B presents a similar issue, where variations in building height combined with shadows introduced substantial interference. While SEIFNet predicted the main change region well, it also produced excessive false positives. Our method maintained strong robustness under these conditions. Sample D illustrates a typical case where trees obscure buildings, causing models to misinterpret the area as containing only vegetation, which resulted in incorrect predictions. ChangeCLIP, AERNet, and DMINet all failed in this scenario. However, our MGCR successfully overcame this challenge, demonstrating strong generalization capabilities when handling complex or occluded situations. As presented in the Table. \ref{tab2}, we compared our model with multiple mainstream methods using four standard evaluation metrics. Our MGCR outperformed the second-best method, ChangeCLIP, by 0.51\% in F1 and 0.92\% in IoU.

	\begin{figure*}[!t]
		\centering
		\includegraphics[width=5.5in]{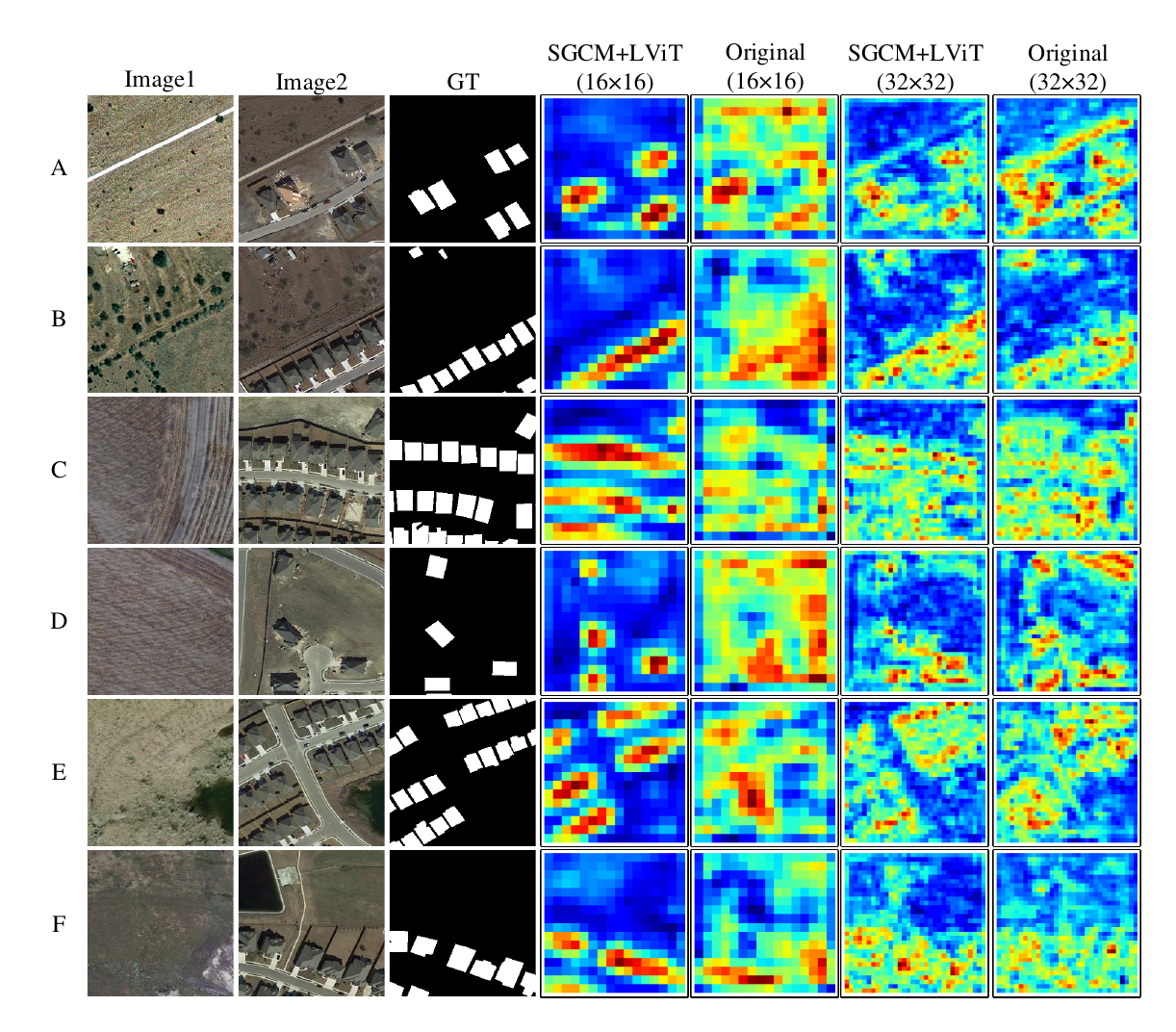}
		\caption{The heatmap visualization results depict the semantic fusion strategies of SGCM and LViT at different pixel locations, compared with the original fusion strategy.}
		\label{fig9}
		\vspace{-10pt}
	\end{figure*}
	
	As shown in the Fig. \ref{fig6}, we selected representative samples A–F from the GZ-CD dataset based on building quantity and density, highlighting different perspectives. Several samples are discussed in detail below. In sample A, while the overall prediction accuracy for the large building area was high, there were noticeable inconsistencies along the building edges compared to the main structure, which affected the prediction results to some extent. Except for ChangeFormer, other mainstream methods produced stripe-like redundant predictions, indicating that our method performs excellently when dealing with edge-related differences. Samples C, D, and E demonstrate the model's performance in detecting changes in complex urban scenes with multi-structured building clusters. Most comparison methods suffered from issues such as buildings being mistakenly merged or individual structures being omitted. In contrast, our method showed strong adaptability and robustness in capturing such structural changes. As shown in the Table. \ref{tab3}, we compared our method with several mainstream models using four commonly used evaluation metrics. Our MGCR achieved the best results, outperforming the second-best method, ChangeCLIP, by 1.17\% in F1 and 1.89\% in IoU.
	
	As shown in the Fig. \ref{fig7}, we selected representative samples A–F from the SYSU-CD dataset, which includes diverse change regions such as building clusters and roads. Below, we provide detailed explanations for selected samples. In sample A, the two temporal images represent vegetation coverage and land use, respectively. Due to the discontinuity in vegetation in the right area, several mainstream methods failed to detect changes completely. Leveraging our novel vision-language alignment and reconstruction strategy, we achieved a complete and accurate prediction of the change area. In sample C, the paired images reflect a change from vegetation to road construction. Competing methods exhibited either large redundant predictions or significant omissions in road detection. Although our method showed some disconnection in the predicted road changes, it effectively captured semantic information, maintaining the correct shape and structural consistency. In sample E, compared to advanced methods like ChangeCLIP and SEIFNet, our method successfully avoided the common tendency of over-predicting large-scale change areas. Instead, it performed precise segmentation without excessive redundancy, demonstrating improved adaptability in complex scenarios. As shown in the Table. \ref{tab4}, we compared our method against several leading approaches using four standard evaluation metrics. Our method outperformed the second-best approach, ChangeCLIP, with an improvement of 0.3\% in F1 and 0.42\% IoU.
	
	Overall, we conducted comprehensive comparison experiments between MGCR and mainstream methods across four public datasets. By selecting representative samples for visual analysis, including various types of change scenarios, and presenting detailed quantitative results in tabular form, we robustly demonstrated the effectiveness of our proposed MGCR network. 
	
	\subsection{Ablation Study}
	To validate the effectiveness of our MGCR method, we conducted ablation studies from multiple perspectives, including adjustments to loss coefficients and the removal of individual modules. Specifically, we set the loss coefficients to 0.8, 0.1, 0.1 and 1, 0, 0 respectively for key parameter sensitivity analysis. As shown in the Table. \ref{tab5}, the values before “/” represent the configuration of 0.8, 0.1, 0.1, while the values after “/” represent the configuration of 1, 0, 0. According to the experimental data, the former demonstrates greater stability and superiority compared to the latter. For the SGCM module, we configured SGCM-V with only visual-conditioned reconstruction and SGCM-L with only language-conditioned reconstruction to investigate the individual effects of visual and language reconstruction. Additionally, we performed ablation on the LViT module. Since removing the LViT module results in the inability to form two branches necessary for loss comparison, it is denoted as a blank “–".

	As shown in the Fig. \ref{fig8}, we selected representative samples from multiple datasets to visualize the ablation experiment results under the loss coefficient setting of 0.8, 0.1, 0.1. In sample A, the extended parts with similar color and texture to the buildings were predicted as change areas by several ablated models. In sample B, MGCR better predicted the small-scale changed buildings and clearly distinguished individual structures, while the ablated models showed varying degrees of missed detections and merged change areas. In sample C, the incomplete SGCM module led to poorer prediction results and a greater tendency for large redundant areas. In sample D, No. 3 visually demonstrates the advantage of multimodal textual data compared to No. 2 and No. 4, avoiding both redundant and missing predictions. Sample F corresponding to No. 5 reflects the overall effect of SGCM, where cross-dimensional alignment of visual and textual features enabled efficient semantic information integration.
	
	As shown in the Fig. \ref{fig9}, to further validate the effectiveness of the mechanisms introduced in MGCR for CD tasks, we conducted a heatmap-based visualization analysis of the model's attention regions during training. At the feature levels of $16\times 16$ and $32\times 32$ pixels, the attention regions of MGCR align closely with the actual change areas, indicating that the model can accurately focus on semantically significant changes. In contrast, models using the original fusion strategy exhibit heatmaps with considerable noise responses, dispersed attention, and more interference. This comparison clearly demonstrates the superior ability of MGCR to focus on change areas and further confirms the critical roles of the SGCM module and the LViT architecture in enhancing the model's semantic understanding and spatial distribution awareness.
	
	
	\section{Discussion}
	This paper leverages LLaVA’s capabilities in visual-language understanding and intelligent dialogue interaction to construct cross-modal textual data based on raw image inputs. By integrating PVT as the visual encoder and CLIP as the text encoder, we build a text-guided cross-modal CD framework. For building CD tasks, the model must first ensure precise localization and segmentation of buildings, followed by extracting difference regions through concatenation and differencing operations. Therefore, we modify prompts to direct LLaVA to generate building-related textual descriptions, while also partially describing other foreground and background information. However, among the large amount of generated text, there remain cases of low image-text similarity. We continue to explore the application of generative AI in multimodal data generation, pushing deeper in building CD tasks to generate textual data with higher image-text matching degrees. At the same time, we aim to broaden the scope in CD tasks to generate textual data that is not limited to buildings but can describe specific and intuitive change targets.
	
	\section{Conclusion}
	In this paper, we proposed a multimodal graph-conditioned vision-language reconstruction network for RSCD. We utilized MLLM to generate cross-modal textual data, which guided the model to achieve multimodal input and multimodal interaction, thereby enhancing the model’s ability to perceive change regions in remote sensing images from multiple perspectives. PVT and CLIP were employed as feature encoders to process image and text information, respectively. Furthermore, we proposed SGCM for the first time to align image-text features, SGCM constructed VL tokens via a graph structure and combined multi-head attention to perform both image and text reconstruction, achieving fully interactive semantic image-text feature information. Subsequently, LViT was used to fuse the reconstructed image-text features, during which the mean squared error between image features and image-text features was calculated to correct feature discrepancies. This, together with binary cross-entropy loss, jointly guided model optimization. Overall, we optimized the strategy for generating textual data via MLLM, processed multimodal features through feature encoders, designed SGCM and leveraged LViT for image-text feature alignment and fusion, and established relevant losses to guide multimodal model training. Comprehensive experiments on multiple public datasets validated the effectiveness of the MGCR multimodal architecture, with results demonstrating that MGCR achieved outstanding performance compared to mainstream methods.
	

	\bibliographystyle{IEEEtran}
	\bibliography{reference}{}
	
\begin{IEEEbiography}[{\includegraphics[width=1in,height=1.25in,clip,keepaspectratio]{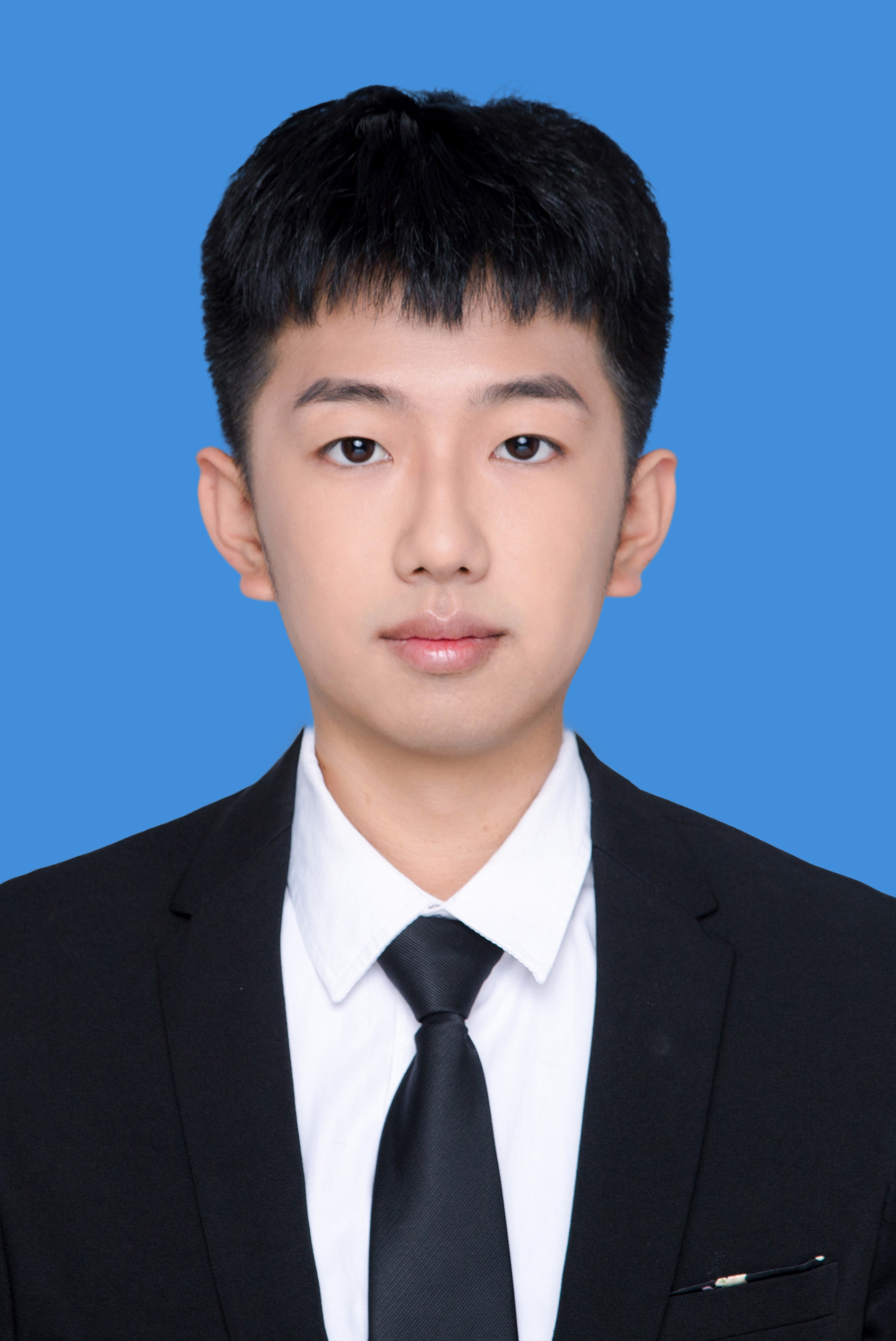}}]{Chengming Wang}
	Chengming Wang received his bachelor’s degree from the School of Computer Science and Technology, Shandong Technology and Business University, Yantai, China in 2023. Currently studying for a master’s degree in the School of Computer Science and Technology, Shandong Technology and Business University, Yantai, Shandong. His research interests include computer graphics, computer vision, and image processing.
\end{IEEEbiography}

\begin{IEEEbiography}[{\includegraphics[width=1in,height=1.25in,clip,keepaspectratio]{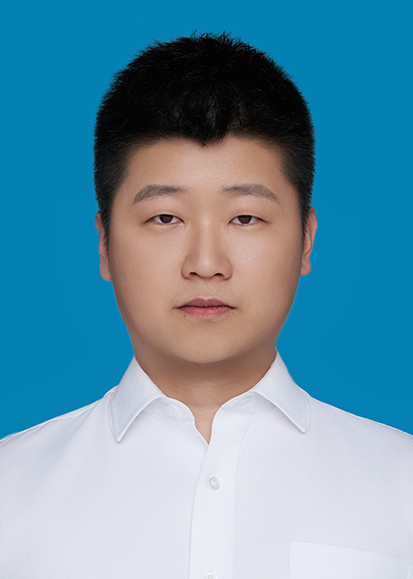}}]{Guodong Fan}
	received a B.Eng. degree in 2018 and a M.Eng. degree in 2021 from Shandong Technology and Business University, and the Ph.D. degree in Software Engineering from Qingdao University, Qingdao, China, in 2025. He is currently a Associate professor at Shandong Technology and Business University.  His research interests include image processing, machine learning, and computer vision.
\end{IEEEbiography}
\begin{IEEEbiography}[{\includegraphics[width=1in,height=1.25in,clip,keepaspectratio]{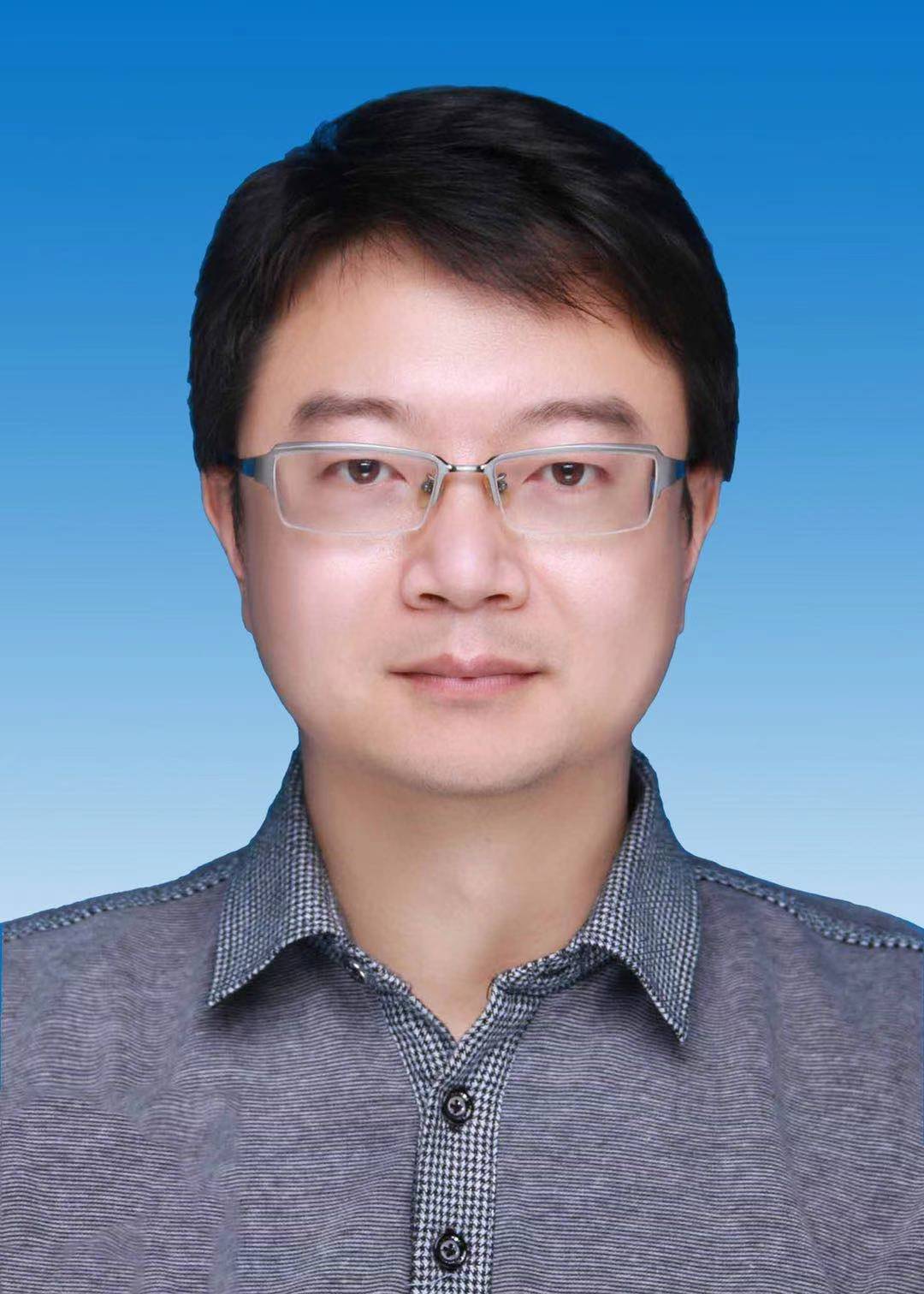}}]{Jinjiang Li}
	received the B. S. and M. S. degrees in computer science from Taiyuan University of Technology, Taiyuan, China, in 2001 and 2004, respectively, the Ph. D. degree in computer science from Shandong University, Jinan, China, in 2010. From 2004 to 2006, he was an assistant research fellow at the institute of computer science and technology of Peking University, Beijing, China. From 2012 to 2014, he was a Post-Doctoral Fellow at Tsinghua University, Beijing, China. He is currently a Professor at the school of computer science and technology, Shandong Technology and Business University. His research interests include image processing, computer graphics, computer vision, and machine learning.
\end{IEEEbiography}

\begin{IEEEbiography}[{\includegraphics[width=1in,height=1.25in,clip]{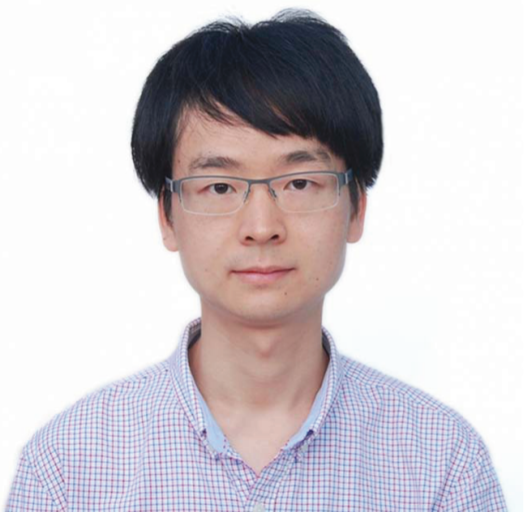}}]%
	{Min Gan} (Senior Member, IEEE) 
	received the B. S. degree in Computer Science and Engineering from Hubei University of Technology, Wuhan, China, in 2004, and the Ph.D. degree in Control Science and Engineering from Central South University, Changsha, China, in 2010. He is currently a professor in the College of Computer Science and Technology, Qingdao University, Qingdao, China. His current research interests include computer vision, statistical learning, system identification and nonlinear time series analysis.  
\end{IEEEbiography}

\begin{IEEEbiography}[{\includegraphics[width=1in,height=1.35in,clip]{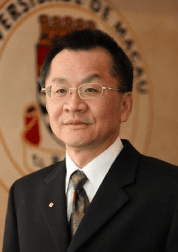}}]%
	{C. L. Philip Chen} (Life Fellow, IEEE) received the M.S. degree in electrical engineering from the University of Michigan, Ann Arbor, MI, USA, in 1985, and the Ph.D. degree in electrical engineering from Purdue University, West Lafayette, IN, USA, in 1988.
	
	He was a Tenured Professor, the Department Head, and an Associate Dean with two different universities in U.S. for 23 years. He is currently the Head of the School of Computer Science and Engineering, South China University of Technology, Guangdong, China. His current research interests include systems, cybernetics, and computational intelligence. He is currently the Dean of the School of Computer Science and Engineering, South China University of Technology, Guangzhou 510641, China. His current research interests include systems, cybernetics, and computational intelligence. 
	
	Dr. Chen received the 2016 Outstanding Electrical and Computer Engineers Award from his alma mater, Purdue University. He was the IEEE SMC Society President from 2012 to 2013 and the Vice President of Chinese Association of
	Automation. He has been the Editor-in-Chief of the IEEE TRANSACTION ON SYSTEMS, MAN, AND CYBERNETICS: SYSTEMS, since 2014 and an associate editor of several IEEE TRANSACTIONS. He was the Chair of TC 9.1 Economic and Business Systems of International Federation of Automatic Control (2015–2017), and also a Program Evaluator of the Accreditation Board of Engineering and Technology Education of the U.S. for Computer Engineering, Electrical Engineering, and Software Engineering Programs. He is a Fellow of AAAS, IAPR, CAA, and HKIE.
\end{IEEEbiography}

\end{document}